\documentclass[12pt,a4in]{article}
\usepackage{graphicx}
\usepackage{geometry}
\usepackage[francais,english]{babel}
\begin{document}
\setcounter{page}{0}
\thispagestyle{empty}

\renewcommand{\thesection}{\Roman{section}}
\begin{center}
\large{\bf Application of  hyperspherical harmonics expansion method to the  low-lying bound S-states of exotic two-muon three-body systems.}\\
\vspace{0.5cm}
\normalsize\rm
MD. ABDUL KHAN\\
\footnotesize
Assistant Professor, Department of Physics, Aliah University, \\ Maulana Abul Kalam Azad Bhavan,\\ DD-45, Sector-I, Salt Lake, Kolkata-700064, India\\ 
{\it Email:} $  drakhan.rsm.phys@gmal.com; drakhan.phys@aliah.ac.in$\\
\end{center}
\vspace{1.0cm}
\rm
\begin{abstract}
Energies of the low-lying bound S-states (L=0) of exotic three-body systems, consisting a nuclear core of charge +Ze (Z being atomic number of the core) and two negatively charged valence muons, have been calculated by hyperspherical harmonics expansion method (HHEM). The three-body Schr\H{o}dinger equation is solved assuming purely Coulomb interaction among the binary pairs of the three-body systems X$^{Z+}\mu^-\mu^-$ for Z=1 to 54. Convergence pattern of the energies have been checked with respect to the increasing number of partial waves $\Lambda_{max}$. For available computer facilities, calculations are feasible up to $\Lambda_{max}=28$ partial waves, however, calculation for still higher partial waves have been achieved through an appropriate extrapolation scheme. The dependence of bound state energies has been checked against increasing nuclear charge Z and finally, the calculated energies have been compared with the ones of the literature.
\end{abstract}
\hspace{1.0cm}{\it Keywords:} Raynal Revai Coefficient, Hyperspherical Harmonics, Hypersph-\\ \hspace*{1.0cm}erical  Harmonics Expansion Method, Renormalized Numerov Method.\\ \hspace*{1.0cm}{\it PACS:} 02.70.-c, 31.15.-Ar, 
31.15.{\cal J}a, 36.10.{\cal E}e.

\section{Introduction}
Exotic few-body systems, consisting electrons, muons, protons, deuteron etc. and their antimatters are becoming more and more significant in atomic spectroscopy, quantum electrodynamics and astrophysics [1-2]. Structural properties of such few-body systems can be investigated in terms of few-body problems involving Coulomb forces. And these few-body Coulomb problems have long research history in non-relativistic quantum mechanics. Starting with few-charged particle systems during the early stages of quantum mechanics [3] such problems pose significant fundamental theoretical and practical importance in atomic-molecular and nuclear-particle physics. As the exotic particles are usually unstable, the atoms (or ions) which they constitute are also very short lived. In practice, atoms of this kind can be formed by stopping accelerated exotic particles in matter. The stopped particles replace one or more electron(s) in an ordinary atom. The first orbit of the exotic particle(s) after capture is very similar in size to that of the electron(s) before ejection. Afterward, it cascades down the ladder of exotic-atom states by x-ray and Auger transitions. If the exotic particle is a negative muon, it passes through various environments before its death in the vicinity of an atomic nucleus [4]. In the early stages it scatters from atom to atom as free electron and gradually gives off its energy until it is captured into an atomic orbit. When it reaches the lowest energy level (1s), it experiences only the Coulomb interaction with the protons in the nucleus and weak interaction with rest of the nucleons. In the case of the hadrons (such as the pion, kaon, or anti-proton) the cascade ends earlier for all exotic atoms except those with atomic number 1 or 2, due to nuclear absorption or annihilation of the particle by the short-range strong interaction. Since exotic particles (except positron) are all much heavier than the electron, they are more strongly bound to the nucleus than electrons, and their transitions during the de-excitation are much more energetic than those of electrons. In addition, exotic particles may come much closer to the nucleus than the electrons in an ordinary atom. The exotic atoms (or ions) are obtained when one or more of the subatomic particles of neutral atom are replaced by one or more exotic particles like muon, pion, kaon, anti-proton etc of the same charge [5]. The most studied exotic few-body Coulomb system are the muonic atoms or ions which are formed by removing one or more orbital electron(s) by one or more negative muon(s). However the present article deals with only those systems where the positively charged nucleus is orbited by two negatively charged- muons. Since the early seventies, muonic atoms are used to measure a number of atomic properties including nature and strength of eletron-muon interaction [6]. In the early fifties, muon was assumed to be very useful probe to measure the electromagnetic properties of nuclei [7]. Magnetic hyperfine structure of muonic atoms was studied by Johnson and Sorensen [8]. Isotopic shifts in muonic spectra of isotopes of Ca, Cr, Cu, Mo etc have been measured by Macagno et al [9]. Krutov and Martynenko studied the hyperfine structure of muonic helium atom ($\mu e ^4_2$He) using perturbation method [10] and also investigated the Lamb-shift in the muonic deuterium ($\mu$D)[11]. Bound-state properties and hyperfine-structure splitting in beryllium-muonic ions are determined by Frolov [12] using highly accurate variational wave functions. Flambaum [13] reported the effect of bound muons, pions, kaons etc on the fission barrier and stability of highly charged nuclei. Many new experiments have been proposed in Muon Science Laboratory, RIKEN[14]. 
Investigation reveals important role of few-body Coulomb system in the cold fusion process (CFP) [15] in which muonic few-body systems experience a strong interplay between nuclear and Coulomb forces involving heavy nuclei such as $(dt\mu)^+$l molecular ion. Furthermore there are large numbers of few-body systems involving antimatter such as antihydrogen ($\overline{H}=\overline{p}e^+$), muonic antihydrogen ($\overline{H}_{\mu}=\overline{p}\mu^+$), antiprotonic helium atom ($\overline{p}+^{4(3)}$He) [16-17] and many more from modern antimatter physics. Studies on such exotic few-body systems involving particle-antiparticle combination may help in checking the CPT law better than usual systems. For example muonic antihydrogen atom ($\overline{H}_\mu$) could be better choice than antihydrogen ($\overline{H}e^+$) for validating CPT law [18-19]. Physics involved in the reaction and dynamics of muonic helium atom has been extensively discussed by T. J. Stuchi et al [20]. In addition to the experiment on the reactions of muonic helium and muonium with H$_2$ by Donald G Fleming and Co-workers [21], several others can also be found in the literature. To study the bound state properties of such exotic system, a number of theoretical methods have been reported in the literature [22-28]. We may refer Rodriguez et al [29] who used angular correlated configuration interaction (ACCI) approach to study the lowest lying states of two-electron and electron-muon three-body atomic systems. Their work includes the energy calculation for negatively charged hydrogen-like systems; neutral helium-like systems, and positively charged lithium like systems. And a  more precise calculation for these systems has been reported by Smith Jr et al [30] and Frolov et al [31-45]. However, first muonic calculations adopting variational approach was started almost 50 years past by Halpern [46], Carter [47-48] and Delves et al [49]. And about 30 years back Vinitsky {\it et al} [50] adopted non-variational approach for the muonic molecular ions. The bound D- state in $dtu$ was 
calculated adopting variational approach by Kamimura in the late eighties [51]. Krivec and Mandelzweig [52] have also reported non-variational calculation for Muonic helium atom ($^4$H$e^{2+}\mu^-e^-$) where they obtained energy up to twelve significant digits employing the Correlated Function HH method. \\The present article deals with only those systems where the positively charged nucleus is surrounded by two negatively charged- muons. The structural properties of such exotic Coulomb systems can be investigated by treating them as a three-body system consisting of a relatively heavy and positively charged nuclear core plus two valence muons. In support of the present model of study we may state the facts that the electromagnetic interaction is much weaker than the strong interaction, hence muon(s) will not perturb the nucleus to any significant degrees and the nuclear degrees of freedom may be neglected as a first approximation. Again, the fact that the muon mass is much smaller than the nuclear mass allows us to regard the nucleus as a almost static source of the Coulomb interaction.  In one of our previous work  [53] we have considered only the ground state of some exotic two-muon atoms and in the present present work we have extended the calculation to some low-lying bound n$^1$S (n=1 to 6) -states of several exotic two-muon three-body systems  hyperspherical harmonics expansion method (HHEM). As discussed in [53], HHEM is a powerful tool for the {\it ab initio} solution of the few-body Schr\"{o}dinger equation, for a given set of potentials of interaction among constituent particles.
Although the method is notorious for its slow convergence particularly for the Coulomb type long range two-body interaction potential, still it is widely used in solving few-body Schr\"{o}dinger equation, for a given set of inter-particle interaction potentials [54-55]. For example, we may refer some of the works of Barnea et al., in which they employed HH for the solution of few-body problems during 1990-2011 [56-60]. And,  the Pisa group of Rosati et al. [61] has used this method to study the binding mechanism of three- and four-nucleon systems viz triton and helium nucleus. The label scheme in this method for a three-body system involves three possible binary interacting pairs which correspond to three different partitions. In the $k^{th}$ partition the particle labeled as $k$ performs the role of a spectator while the remaining two labeled as $i$ and $j$ form the interacting pair. For the calculation of matrix element of V($r_{ij}$), the interaction potential of the $(ij)$ pair, it is then convenient to expand the chosen HH in the set of HH corresponding to the partition in which $\vec{r_{ij}}$ is proportional to the first Jacobi vector [62].
 To do this we need the transformation coefficients, called Raynal-Revai coefficients (RRC), from one choice of partition to another. Raynal and Revai [63] obtained an expression for these coefficients for a three-body system containing particles of arbitrary masses. In this work, RRC [62-64] has been used in the numerical computation of potential matrix elements of the two-body interactions, involved in the three-body systems consisting of a positively charged nucleus plus two negatively charged muons. The energies calculated for the low-lying bound S-states have been compared with the ones of the literature, dependence of convergence pattern of energies have been checked against increasing i) nuclear charge Z and ii) number of partial waves $\Lambda_m$ included in the calculation.
In Section II, we briefly describe hyperspherical harmonics expansion method and the transformation coefficients between two sets of HH belonging  to two different partitions. In Section III, we shall briefly discuss the use of RRC in the calculation of energies for the low-lying bound S-states of helium-like systems consisting of a positively charged core plus two valence muons. The results of calculated observables will be compared to the ones of the literature wherever available.

\section{HHE Method}
For a general three-body system consisting particles of unequal masses $m_{i}$, $m_{j}$, $m_{k}$, the label scheme is  shown in Fig.1.

\begin{figure}
\centering
\fbox{\includegraphics[width=0.75\linewidth, height=0.6\linewidth]{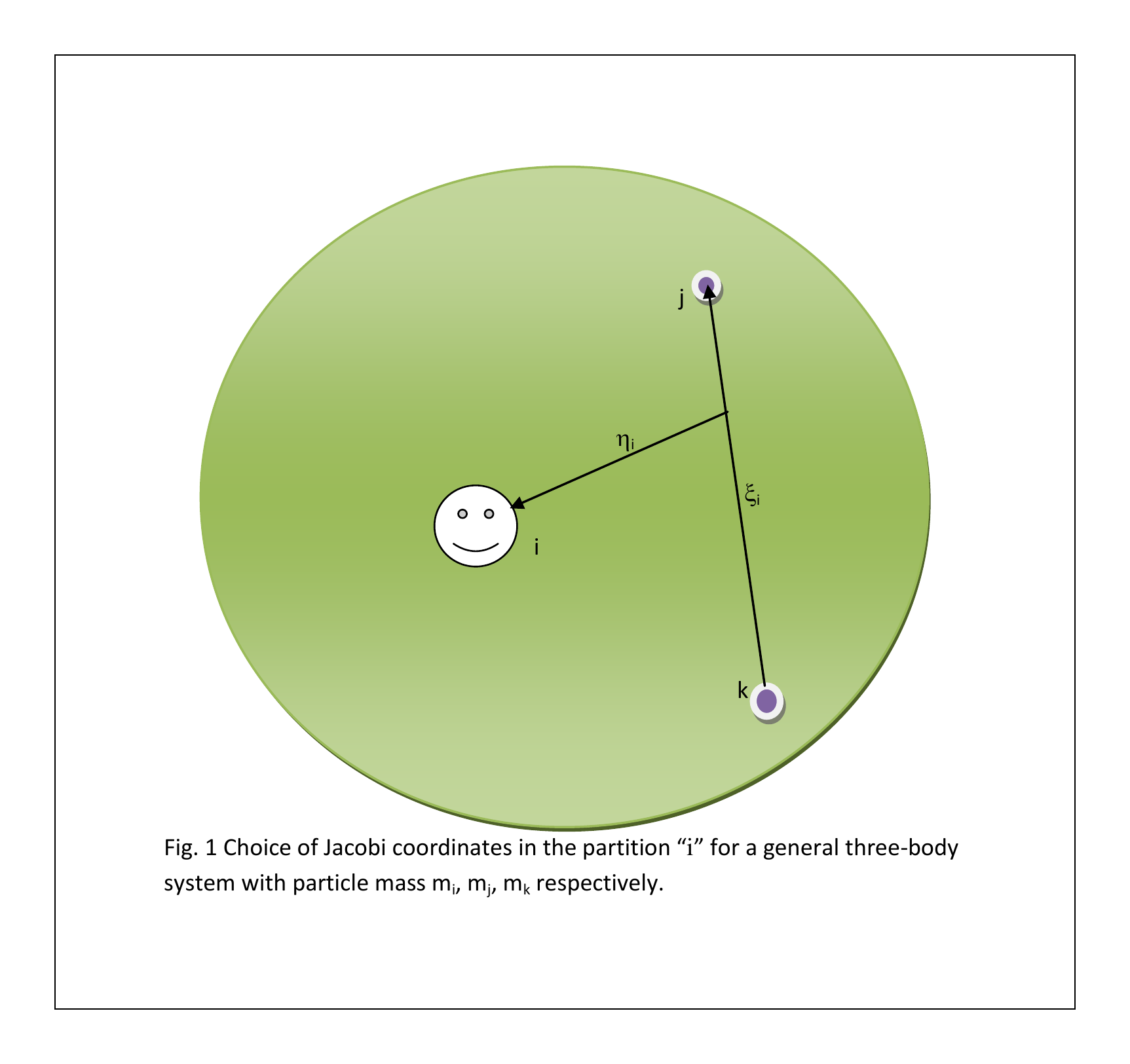}}
\caption{Label scheme for general three-body system and choice of Jacobi coordinates in the partition $\lq\lq i$".}
\label{fig:boxed_graphic}
\end{figure}
The Jacobi coordinates [69] to describe the relative motion in the partition - $\lq\lq i$" are defined as: 
\begin{equation}
 \left.  \begin{array}{ccl}
   \vec{\xi_{i}} & = & \left[ \frac{m_{j} m_{k}M}{m_{i}(m_{j}+m_{k})^{2}}
\right] 
^{\frac{1}{4}} (\vec{r_{j}} - \vec{r_{k}}) \\ 
   \vec{\eta_{i}} & = & \left[ \frac{m_{i} (m_{j}+m_{k})^{2}}
{m_{j} m_{k} M}\right]^{\frac{1}{4}} \left( \vec{r_{i}} - \frac{m_{j}
\vec{r_{j}} + m_{k} \vec{r_{k}}}{ m_{j} + m_{k}} \right) 
    \end{array}  \right\} 
\end{equation}
where $M=m_i+m_j+m_k$ and the sign of $\vec{\xi_{i}}$ is determined by the condition that ($ i, j, k $) should form a cyclic permutation of (1, 2, 3).\\
Eq.(1) represents set of Jacobi coordinates which corresponds to the partition, in which, the particle labeled $\lq\lq i$" is the spectator and particles labeled $\lq\lq j$" and $\lq\lq k$" form the interacting pair.
In terms of the  hyperspherical variables [53]
\begin{equation}
 \left. \begin{array}{cclcccl}
 \xi_{i} & = & \rho \cos \phi_{i}&;& \eta_{i} & = & \rho \sin \phi_{i}\\
\rho&=& \sqrt{ \xi_{i}^{2} + \eta_{i}^{2}}&;&\phi_i&=&tan^{-1}(\eta_i/\xi_i)
	  \end{array} \right\} 
\end{equation}
the three-body Schr\H{o}dinger's equation appears as
\begin{equation}
\left[ - \frac{\hbar^{2}}{2\mu}\left\{ \frac{\partial^2}{\partial\rho^2}+ \frac{5}{\rho} \frac{\partial}{\partial\rho}+
\frac{\hat{{\Lambda}}^{2}(\Omega_{i})}{\rho^{2}} \right\}+ V (\rho, \Omega_{i}
) - E \right] 
\Xi (\rho, \Omega_{i} ) \:=\: 0
\end{equation}
in which $ \Omega_{i} \rightarrow \{ \phi_{i}, \theta_{\xi_{i}}, \phi_{\xi_{i}},
\theta_{\eta_{i}}, \phi_{\eta_{i}} \}$, 
${\mu\: =\: \left[ \frac{m_{i} m_{j} m_{k}}{M} \right]^{\frac{1}{2}}}$ is an effective
 mass parameter, $V (\rho, \Omega_{i})$ = $ V_{jk} + V_{ki} + V_{ij} $ is the
total  interaction potential, and  square of hyper angular momentum operator $\hat{\Lambda}^{2}(\Omega_{i})$ satisfies the eigenvalue equation [53]
\begin{equation}
\hat{{\Lambda}}^{2}(\Omega_{i}) {\Theta}_{\Lambda \alpha_{i}}(\Omega_{i})=\Lambda
(\Lambda + 4 )  
{\Theta}_{\Lambda \alpha_{i}}(\Omega_{i})
\end{equation}
where the eigenfunction $\Theta_{\Lambda \alpha_{i}}(\Omega_{i})$ is called the hyperspherical harmonics (HH). The normalized HH with specified three-body total orbital angular momentum $L(=\mid\vec{l_{\xi_{i}}} + \vec{l_{\eta_{i}}}\mid)$ and its projection $M$
 is given by
\begin{equation}
 \begin{array}{rcl}
{\Theta}_{\Lambda \alpha_{i}}( \Omega_{i} )& \equiv &
{\Theta}_{\Lambda l_{\xi_{i}} l_{\eta_{i}} L M}(\phi_{i}, \theta_{\xi_{i}},
\phi_{\xi_{i}}, \theta_{\eta_{i}}, \phi_{\eta_{i}})\\
 & \equiv & ^{(2)}P_{\Lambda}^{l_{\xi_{i}} l_{\eta_{i}}}(\phi_{i})
\left[ Y_{l_{\xi_{i}} m_{\xi_{i}}}(\theta_{\xi_{i}}, \phi_{\xi_{i}}) Y_{l_{\eta_{i}}
m_{\eta_{i}}}(\theta_{\eta_{i}}, \phi_{\eta_{i}}) \right]_{L M}
 \end{array}
\end{equation}
where $\alpha_{i}$ represents $\{ l_{\xi_{i}}, l_{\eta_{i}}, L, M \}$ and $[ ]_{L M}$ stands for coupling of angular momentum.
The  hyper-angular momentum quantum number $\Lambda=2n_i+ l_{\xi_{i}} + l_{\eta_{i}}, \:n_i$ being non-negative integer is not a good quantum number for the three-body system.\newline
For a given partition (say partition $\lq\lq i$"), the  wave-function $\Xi(\rho, \Omega_{i})$  is expanded in the complete 
set of HH 
\begin{equation}
\Xi(\rho, \Omega_{i}) = \sum_{\Lambda\alpha_{i}}\rho^{-5/2}F_{\Lambda\alpha_{i}}   
  (\rho) {\Theta}_{\Lambda\alpha_{i}}(\Omega_{i})
\end{equation}
 Insertion of eq.(6) in eq.(3), use of  eq.(4) and application of the ortho-normality of HH, leads to the set of coupled
differential equations (CDE) in $\rho$
\begin{equation}
\begin{array}{cl}
& \left[ -\frac{\hbar^{2}}{2\mu} \frac{d^{2}}{d\rho^{2}}
+\frac{(\Lambda+3/2)(\Lambda+5/2) \hbar^2}{2\mu\rho^{2}} - E \right] 
F_{\Lambda \alpha_{i}}(\rho)  \\
+ & \sum_{\Lambda^{\prime} \alpha_{i}^{\prime}} <\Lambda \alpha_{i}
\mid V(\rho, \Omega_{i}) \mid \Lambda^{\prime} \alpha_{i}^{\prime}
>F_{\Lambda^{\prime} \alpha_{i}^{~\prime}}(\rho) \: = \: 0.
\end{array}
\end{equation}
where
\begin{equation}
<\Lambda \alpha_{i} |V(\rho,\Omega_i)| \Lambda^{\prime}, \alpha_{i}^{\prime}> = \int
{\Theta}_{\Lambda\alpha_{i}}^{*}(\Omega_{i}) V(\rho, \Omega_{i}) {\cal
Y}_{\Lambda^{\prime} 
 \alpha_{i}^{~\prime}}(\Omega_{i}) d\Omega_{i}
\end{equation}
   For central interaction potentials, calculation of the matrix elements of the form $<{\Theta}_{\Lambda \alpha_{i}}(\Omega_{i}) \mid V_{j k}(\xi_{i}) \mid {\cal Y}_{\Lambda^{\prime}  \alpha_{i}^{\prime}}(\Omega_{i})>$,  in the partition $\lq\lq i"$, is straight forward, while the same becomes very complicated for $<{\Theta}_{\Lambda \alpha_{i}}(\Omega_{i}) \mid V_{i j}(\xi_{k}) \mid {\Theta}_{\Lambda^{\prime}  \alpha_{i}^{~\prime}}(\Omega_{i})>$ or $<{\Theta}_{\Lambda \alpha_{i}}(\Omega_{i}) \mid V_{k i} (\xi_{j})|{\Theta}_{\Lambda^{\prime} \alpha_{i}^{~\prime}}(\Omega_{i})>$ even for central potentials, since $\xi_{k}$ or $\xi_{j}$ involves the polar angles $\hat{\xi_{i}}$ and $\hat{\eta_{i}}$. From eq.(1), we may write \begin{equation}
 \left. \begin{array}{rcl}
\vec{\xi_{k}} & = & - \cos \sigma_{ki} \vec{\xi_{i}} + \sin \sigma_{ki}
\vec{\eta_{i}}\\ 
\vec{\eta_{k}} & = & - \sin \sigma_{ki} \vec{\xi_{i}} - \cos \sigma_{ki} \vec{\eta_{i}}
	  \end{array} \right\}
\end{equation}
where $\sigma_{ki}$ = $\tan^{-1} \{(-1)^{P} \sqrt{\frac{M m_{j}}{m_{i} m_{k}}}
\}$, P being odd (even) if ($kij$) is an odd (even) permutation of the
triad (1 2 3).\newline
For any arbitrary shape of the central potential with non-zero L,  calculation becomes 
inaccurate and slow, since most of the five dimensional integrals have to be 
done numerically. However, evaluation of the latter matrix elements can be greatly simplified in the way it is described in                                                                                                                                                                                                                                                                                                                                                                                                                     [53].\newline
As the complete sets of HH functions $\{{\Theta}_{\Lambda \alpha_{i}}(\Omega_{i})\}$, $\{{\Theta}_{\Lambda
\alpha_{j}}(\Omega_{j})\}$ or $\{{\Theta}_{\Lambda \alpha_{k}}(\Omega_{k})\}$ span the same five dimensional angular
hyperspace, any particular member of a given set, say ${\Theta}_{\Lambda \alpha_{i}}(\Omega_{i})$ can be expanded in the complete set of $\{{\Theta}_{\Lambda \alpha_{j}}(\Omega_{j})\}$ through a unitary transformation:
\begin{equation}
{\Theta}_{\Lambda \alpha_{i}}(\Omega_{i})=\sum_{\alpha_{j}} < \alpha_{j} \mid
 \alpha_{i} >_{\Lambda L} {\Theta}_{\Lambda \alpha_{j}}(\Omega_{j}) 
\end{equation}
Again, since $\Lambda, L, M$ are conserved for eq.(10) and there is rotational degeneracy 
with respect to the quantum number $M$ for spin independent forces, we have 
\begin{equation}
<\alpha_{j} \mid \alpha_{i}>_{\Lambda L} = <l_{\xi_{j}} l_{\eta_{j}} \mid l_{\xi_{i}}
l_{\eta_{i}}>_{\Lambda L}
\end{equation}
Thus, we can rewrite eq(10) as 
\begin{equation}
{\Theta}_{\Lambda\alpha_{i }}(\Omega_{i}) = \sum_{l_{\xi_{j}} l_{\eta_{j}}}<l_{\xi_{j}}
l_{\eta_{j}} \mid  l_{\xi_{i}} l_{\eta_{i}}>_{\Lambda L} {\Theta}_{\Lambda\alpha_{j}}(\Omega_{j})
\end{equation}
 The M independent coefficients involved in eq(11) and (12) are called the Raynal-Revai Coefficients (RRC). Using these coefficients, 
 the matrix element of a central interaction $V_{ij}$ becomes 
\begin{equation}
 \begin{array}{rcl}
<{\Theta}_{\Lambda \alpha_{i}}(\Omega_{i}) \mid V_{ij} (\xi_{k}) \mid {\Theta}_{\Lambda^{\prime} 
 \alpha_{i}^{~\prime}}(\Omega_{i})> & = & \sum_{l_{\xi_{k}}^{~\prime}
l_{\eta_{k}}^{~\prime} l_{\xi_{k}} l_{\eta_{k}}}<l_{\xi_{k}} 
l_{\eta_{k}} \mid  l_{\xi_{i}} l_{\eta_{i}}>_{\Lambda L}^{*}\\
&\times & <l_{\xi_{k}}^{~\prime}
l_{\eta_{k}}^{~\prime} \mid l_{\xi_{i}}^{~\prime} l_{\eta_{i}}^{~\prime}
>_{\Lambda^{\prime} L} \\  &\times &  <{\Theta}_{\Lambda
\alpha_{k}}(\Omega_{k})\mid V_{ij}(\xi_{k}) \mid {\Theta}_{\Lambda^{\prime}
 \alpha_{k}^{~\prime}}(\Omega_{k})>
 \end{array}
\end{equation}
 The matrix element on the right side of eq.(13) resembles the matrix element of $V_{jk}$ in the partition $\lq\lq i$" and can be calculated in a simple manner. Thus, one can calculate matrix element of $V_{ij}$ easily by computing RRC's involved in eq.(13) 
using their elaborate expressions from [62-64]. Similar treatment can be applied for the calculation of the matrix element of $V_{ki}$. At this point we may also refer the analytical calculation of matrix elements of the effective potential in correlation function HH method by Krivec and Mandelzweig [65].

\section{Application to Coulomb three-body problem: two-muon atoms and ions}
We apply HHEM together with the idea of Raynal-Revai Coefficients to the bound S-states of three-body Coulomb systems consisting a positively charged nucleus of arbitrary Z ($\leq$ 54) plus two negatively charged muons ($\mu^-$). We assign the label $\lq\lq i$" to the nucleus (of mass $m_N$ and charge +Ze), and labels $\lq\lq j$" and $\lq\lq k$" to two muons (of mass $m_j =m_k = m$  and charge -e) respectively. Jacobi coordinates of eq(1) in the partition $\lq\lq i$", for this particular choice of masses becomes 
\begin{equation}
 \left. \begin{array}{rcl}
  \vec{\xi_{i}} & = &\beta_{i} (\vec{r_{j}} - \vec{r_{k}}) \\
  \vec{\eta_{i}} & = &\frac{1}{\beta_{i}} (r_{i} - \frac{\vec{r_{j}}+
\vec{r_{k}}}{2}) 
	  \end{array} \right\}
\end{equation}
where we relate $\beta_{i}$ =
$\left[ \frac{m_{N}+2 m}{4 m_{N}} \right]^{\frac{1}{4}}$ to the system effective mass $\mu$ in the way  
\begin{equation}
 \begin{array}{cclcl}
\mu & = & m \left( \frac{m_{N}}{m_{N}+2m} \right)^{\frac{1}{2}} & = &
\frac{m}{2\beta_{i}^{2}} 
 \end{array} 
\end{equation}
In muon atomic units (i.e., $\hbar^{2}$=$m_{\mu}$=$m$=$e^{2}$=1) eq(7) takes the form  
\begin{equation}
\begin{array}{lcl}
 \left[ -\beta_i^2 \left\{ \frac{d^{2}}{d\rho^{2}}
-\frac{(\Lambda+3/2)(\Lambda+5/2)}{\rho^{2}}\right\} - E \right] 
F_{\Lambda \alpha_{i}}(\rho)&&  \\
+ \sum_{\Lambda^{\prime} \alpha_{i}^{~\prime}} < \Lambda \alpha_{i}
\mid     \frac{\gamma_{i}}{\rho 
cos \phi_{i}} - \frac{Z}{\rho\left|\gamma_{i}sin \phi_{i}~
\hat{\eta_{i}}-
\frac{1}{2\gamma_{i}} cos\phi_{i}\hat{\xi_{i}}\right|}&& \\
  -\frac{Z}{\rho\left|
\gamma_{i}sin \phi_{i} \hat{\eta_{i}}+\frac{1}{2\gamma_{i}}cos\phi_{i}
\hat{\xi_{i}}\right|}
             \mid  \Lambda^{\prime} \alpha_{i}^{~\prime}
>F_{\Lambda^{\prime} \alpha_{i}^{~\prime}}(\rho)& = & 0
\end{array}
\end{equation}
The mass of the particles involved in this work can be found in [53,66-67] and the energies presented in [1] in atomic unit (a.u.) have been converted to muon atomic unit (m.a.u.) following the conversion relation $ 1\: m.a.u. =206.7682838\: a.u.$, taking muon mass $m_{\mu}=206.7682838 m_e$ provided by Mohr and Taylor [68].
Calculation of potential matrix elements of muon-nucleus Coulomb interactions $V_{ij}$ and $V_{ki}$ in the partition $\lq\lq i$" are greatly simplified by the use of RRC, following- prescriptions of previous section and the same described in our previous work [53]. For two-muon three-body systems, 
\begin{equation}
 \begin{array}{ccccl}
\beta_{j}& = &\beta_{k}& = &\left[ 1 - \frac{m^{2}}{(m_{N} +m)^{2}}
\right]^{\frac{1}{4}}  
 \end{array}
\end{equation}
and for heavy nucleus, ${m_{N} \gg m}$, $\mu\approx m$, ${\gamma_{i} \approx
\frac{1}{\sqrt{2}}}$, ${\beta_{j}=\beta_{k} \simeq}$ 1.\\  In eq(6), we expand the three-body relative wave function in the complete set HH appropriate to the partition $\lq\lq i$". For the bound S-states of two-muon systems, the total orbital angular momentum, $L$=0 and the spin part of the total wave function for two-muons  is anti-symmetric. Again, since $L=0$, ${l_{\xi_{i}}= l_{\eta_{i}}}$. So, the set of quantum numbers represented by $\alpha_{i}$ is ${\left\{ l_{\xi_{i}}, l_{\xi_{i}}, 0, 0 \right\}}$. Thus, the quantum numbers ${\left\{\Lambda\alpha_{i} \right\}}$ can be represented by ${\left\{\Lambda l_{\xi_{i}}\right\}}$ only. Furthermore, since the space part of the wave function must be symmetric, under the exchange of the two muons, only even
values of $l_{\xi_{i}}$ ($ \leq \Lambda/2$) are allowed. Corresponding HH can then be written as
\begin{equation}
 \begin{array}{ccl}
{\Theta}_{\Lambda \alpha_{i}}(\Omega_{i}) & \equiv & {\Theta}_{\Lambda l_{\xi_{i}}
l_{\xi_{i}} 0 0}(\Omega_{i}) \\ 
 & =  & ^{(2)}P_{\Lambda}^{l_{\xi_{i}} l_{\xi_{i}}}(\phi_{i})\left[Y_{l_{\xi_{i}}
m_{\xi_{i}}}(\theta_{\xi_{i}},\phi_{\xi_{i}}) 
Y_{l_{\xi_{i}} - m_{\xi_{i}}} (\theta_{\xi_{i}},\phi_{\xi_{i}})\right]_{0 0} \\
 &  & \left( \Lambda \: even \: and \: l_{\xi_{i}} \: = \: 0, 2, 4, \ldots ,\Lambda/2
\right) . 
 \end{array}
\end{equation}
 The matrix element of the muon-muon repulsion term in our chosen partition $\lq\lq i$", is 
\begin{equation}
  \begin{array} {rcl}
    <\Lambda^{\prime}l_{\xi_{i}}^{\prime}|\frac{\gamma_{i}}{\rho ~cos\phi_{i}}|\Lambda
l_{\xi_{i}}>&=&\frac{\gamma_{i}}{\rho} \delta_{l_{\xi_{i}}^{~\prime}, 
 l_{\xi_{i}}} \int_{0}^{\pi /2} { ^{(2)}P_{\Lambda^{\prime}}}^{l_{\xi_{i}}
l_{\xi_{i}}}(\phi) \\
&&\times {^{(2)}P_{\Lambda}}^{l_{\xi_{i}} l_{\xi_{i}}}(\phi) \sin^{2}\phi ~\cos~\phi ~d\phi\\
  \end{array}
\end{equation}
in which the suffix $i$ on $\phi$ has been dropped deliberately, since $\phi$ is only a variable of integration. In the same way, the matrix element of the third term (i.e. the muon-nucleus attraction term) in
the partition $\lq\lq k$" is \begin{equation}
  \begin{array} {rcl}
<\Lambda^{\prime} l_{\xi_{k}}^{\prime} \mid \frac{\gamma_{k}}{\rho cos\phi_{k}}
\mid \Lambda l_{\xi_{k}}>&=& \frac{\gamma_{k}}{\rho} \delta_{l_{\xi_{k}}^{\prime},
l_{\xi_{k}}} \int_{0}^{\pi /2}  {^{(2)}P_{\Lambda^{\prime}}}^{l_{\xi_{k}}
l_{\xi_{k}}}(\phi) \\
&&\times {^{(2)}P_{\Lambda}}^{l_{\xi_{k}} l_{\xi_{k}}}(\phi) \sin^{2}\phi \cos~\phi ~d\phi\\
  \end{array}
\end{equation}
 An identical relation holds for the matrix element of the last term of eq.(16) in
the partition $\lq\lq j$" [53]. Eqs.(19) and (20) show that the matrix elements are
essentially the same in their respective partitions, although $l_{\xi_{k}}$
and $l_{\xi_{j}}$ are not restricted to only even integer values. Each 
 involves only a single, one dimensional integral to be performed
numerically. Using eq.(13), matrix elements of the third and fourth terms
of eq. (16) in our chosen partition (i.e., partition $\lq\lq i$") become
\begin{equation}
 \begin{array}{rcl}
<\Lambda^{\prime} l_{\xi_{i}}^{\prime}\mid\frac{Z}{r_{ij}}\mid \Lambda
l_{\xi_{i}}> & = & \sum_{l_{\xi_{k}}}<l_{\xi_{k}}l_{\xi_{k}} \mid
l_{\xi_{i}}^{\prime}l_{\xi_{i}}^{\prime}>_{\Lambda^{\prime}0}^{\*} 
<l_{\xi_{k}} l_{\xi_{k}}\mid l_{\xi_{i}} l_{\xi_{i}}>_{\Lambda  
0} \\
 && <\Lambda^{\prime} l_{\xi_{k}}\mid\frac{Z \gamma_{k}}{\rho
cos\phi_{k}}\mid \Lambda l_{\xi_{k}}>. 
 \end{array}
\end{equation}
and
\begin{equation}
 \begin{array}{rcl}
<\Lambda^{\prime}l_{\xi_{i}}^{\prime}\mid\frac{Z}{r_{ik}}\mid \Lambda l_{\xi_{i}}>
&=&\sum_{l_{\xi_{j}}} <l_{\xi_{j}}l_{\xi_{j}}\mid 
 l_{\xi_{i}}^{~\prime}l_{\xi_{i}}^{\prime}>_{\Lambda^{\prime}0}^{\*}
<l_{\xi_{j}}l_{\xi_{j}}\mid 
l_{\xi_{i}}l_{\xi_{i}}>_{\Lambda 0} \\ 
 &&<\Lambda^{\prime}l_{\xi_{j}}\mid\frac{Z\gamma_{j}}{\rho
cos\phi_{j}}\mid \Lambda l_{\xi_{j}}>.  
 \end{array}
\end{equation}
Sums over $l_{\xi_{k}}^{\prime}$ and
$l_{\xi_{j}}^{\prime}$ respectively in eqs.(21) and (22) have been performed using the Kronecker
 - $\delta$'s in eq.(20) and a similar one with suffix $k$ replaced by suffix
 $j$. Thus the evaluation of the matrix elements of $\underline{all}$ the
potential components become practically simple and easy to handle numerically.
One of the major drawback of HH expansion method is the slow rate of convergence for long range Coulomb-type interaction potentials, although the rate of convergence for short-range interaction potentials is reasonably fast [69-70]. So, to reach expected  degree of convergence, sufficiently large $\Lambda_m$ value is to be incorporated in the calculation. But, if all $\Lambda$ values up to a maximum of $\Lambda_m$ are included in the HH expansion then the number (K) of such basis state function will be given by 
\begin{equation}
 K=\left\{ \begin{array}{l}
  (\frac{\Lambda_m}{4}+1)^2 \: \:if \: \frac{\Lambda_m}{2} \: is\: even\\
  \frac{(\frac{\Lambda_m}{2}+1)(\frac{\Lambda_m}{2}+3)}{4} \: \:if\: \frac{\Lambda_m}{2} \: is \: odd. 
	  \end{array} \right.
\end{equation}
From eq.(23), one can easily note that the total number of basis states and hence the size of coupled differential equations (CDE) [eq.(7)] increases rapidly with increase in $\Lambda_m$. For instance, one has to solve 625 CDE for $\Lambda_m=96$ which leads the computation towards instability. 
The present calculation is performed on a core-i3 based desktop computer which allowed us to solve up to $\Lambda_m=28$ reliably. The calculated bound state energies ($B_{\Lambda_m}$) for values of $\Lambda_m$ up to 28 are presented in columns 2 - 12 of Table-1 for some low-lying bound S-states of two-muon three-body systems with nucleus of arbitrary charge Z like-$^{\infty}$He$^{2+}\mu^-\mu^-$,  $^{\infty}$B$e^{4+}\mu^-\mu^-$, $^{\infty}$C$^{6+}\mu^-\mu^-$, $^{\infty}$Ne$^{10+}\mu^-\mu^-$ and $^{\infty}$Ge$^{32+}\mu^-\mu^-$. The energies for higher $\Lambda_m$ values are estimated following an extrapolation scheme prescribed  by T. R. Schneider [71] described in great details in [53]. The extrapolated values are presented in column 4 of Table-3 and columns 4 and 8 of Table-4 respectively. In columns 5 \& 6 of Table-3  results of some other calculations for the same states wherever available have been listed for comparison with our values
The pattern of convergence in the energy of the low-lying bound S-states with respect to increasing $\Lambda_m$ can be checked by gradually increasing $\Lambda_m$ values in suitable steps ($d\Lambda$) and comparing the energy difference $\Delta B=B(\Lambda+d\Lambda)-B(\Lambda)$ with that found in the previous step. From the calculated energy data recorded in Table-1, it can be seen that for $\Lambda_m=28$, energy of the lowest (n=1) bound S-state of He$^{2+}\mu^-\mu^-$ converges up to 3rd decimal places while that of the excited (n=5) bound S-state converged only up to the 1st decimal places. Convergence trend in the remaining cases also follow the same pattern. Thus,  energy of the lowest-lying bound S-states converges faster than the energy of  higher excited S-states. Again, for increasing $\Lambda_m$,  energy of any particular low-lying bound S-state in a relatively  lighter system converges faster than the energy of the corresponding state in heavier system. In addition to the above, we may also state  that the energy of any particular bound S-state of two-muon system of  lower nuclear charge Z, converges faster than the energy of the corresponding bound S-state of two-muon systems of relatively greater nuclear charge Z. For justification of the forgoing remarks we first estimated the difference in energy $\Delta B=B(\Lambda_m=28)-B(\Lambda_m=24)$ using energy values recorded in columns 2 and 3 of Table-1 for the 1$^1$S and 5$^1$S states respectively of He$^{2+}\mu^-\mu^-$ and compared them. These estimates are 0.0020 mau and 0.0569 mau respectively. Similar results can also be seen by a simple eye estimation from Fig. 5  drawn for Ge$^{32+}\mu^-\mu^-$. And then we estimated $\Delta B=B(\Lambda_m=28)-B(\Lambda_m=24)$ using data presented in columns 2, 4, 6, 7, 8 of Table-1 for the 1$^1$S state of He$^{2+}\mu^-\mu^-$, Be$^{4+}\mu^-\mu^-$, C$^{6+}\mu^-\mu^-$, Ne$^{10+}\mu^-\mu^-$ and Ge$^{32+}\mu^-\mu^-$ respectively and compared them. These estimates are 0.0020, 0.0062, 0.0128, 0.0336 and 0.3131 m.a.u. respectively. The same results can also be seen from Fig.3 obtained for the 5$^1$S state of- He$^{2+}\mu^-\mu^-$ (Z=2), Be$^{4+}\mu^-\mu^-$ (Z=4), C$^{6+}\mu^-\mu^-$ (Z=6) and Ne$^{10+}\mu^-\mu^-$(Z=10) respectively. In this way we justify our forgoing remarks.
Furthermore, it could also be mentioned here that, although, the direct computation of the matrix element of  ${\frac{1}{r_{ij}}}$ in the partition $\lq\lq i$" is possible by the method of ref.[61 of 53], it is not possible for potentials other than Coulomb or harmonic type. For an arbitrary shape of interaction potential, a direct computation of the matrix element of the potential will involve five dimensional angular integrations which lead the calculation very time consuming and leaves doors open for inaccuracies to creep in easily. Thus for accurate and faster computation of energy role of RRC in HH method is unique and essential .
\begin{figure}
\centering
\fbox{\includegraphics[width=0.75\linewidth, height=0.6\linewidth]{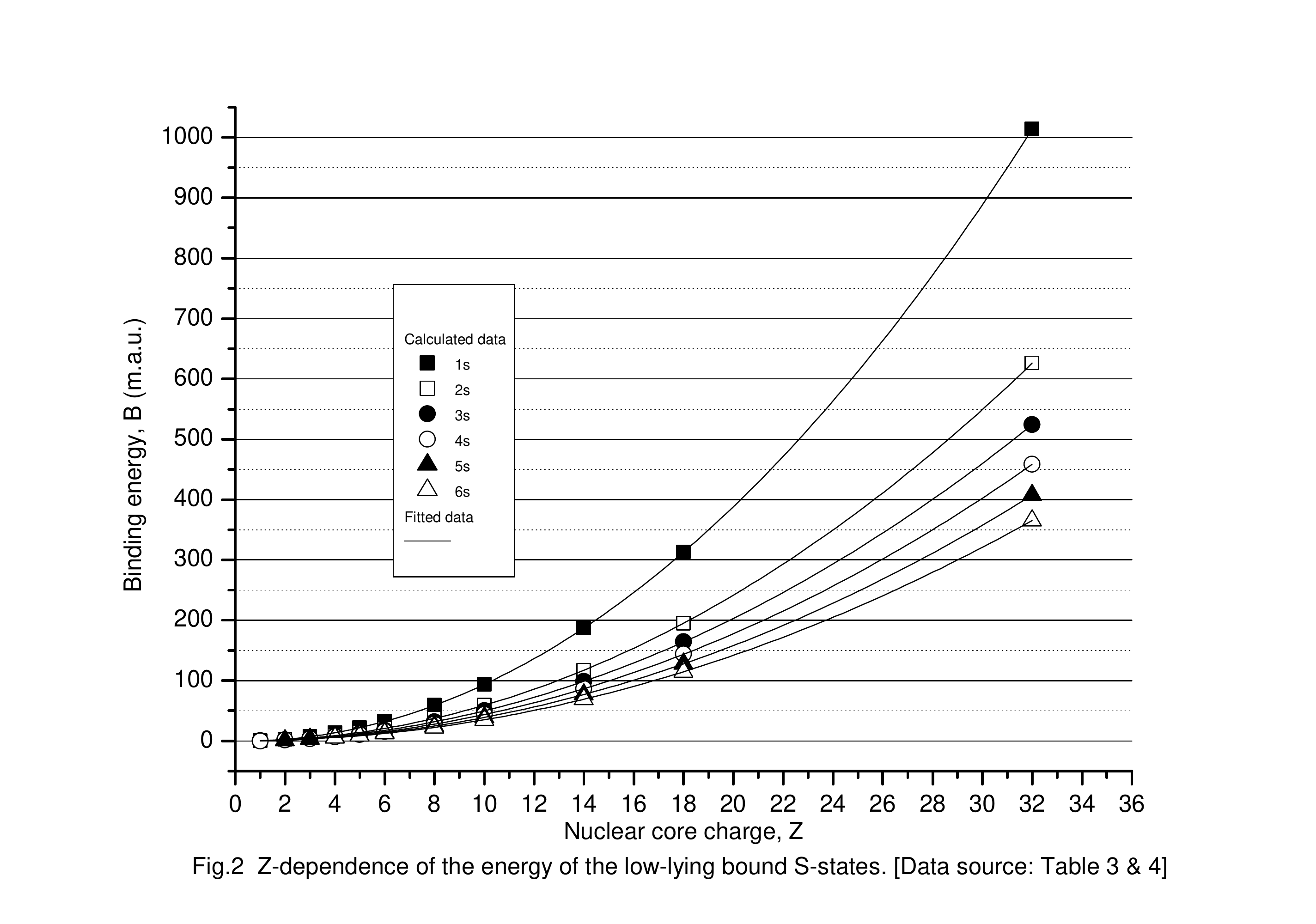}}
\caption{ Dependence of the energy (B) of the low-lying bound S-states of two-muon three-body systems on the increase in nuclear charge Z.}
\label{fig:boxed_graphic}
\end{figure}
\newpage
\begin{figure}
\centering
\fbox{\includegraphics[width=0.75\linewidth, height=0.6\linewidth]{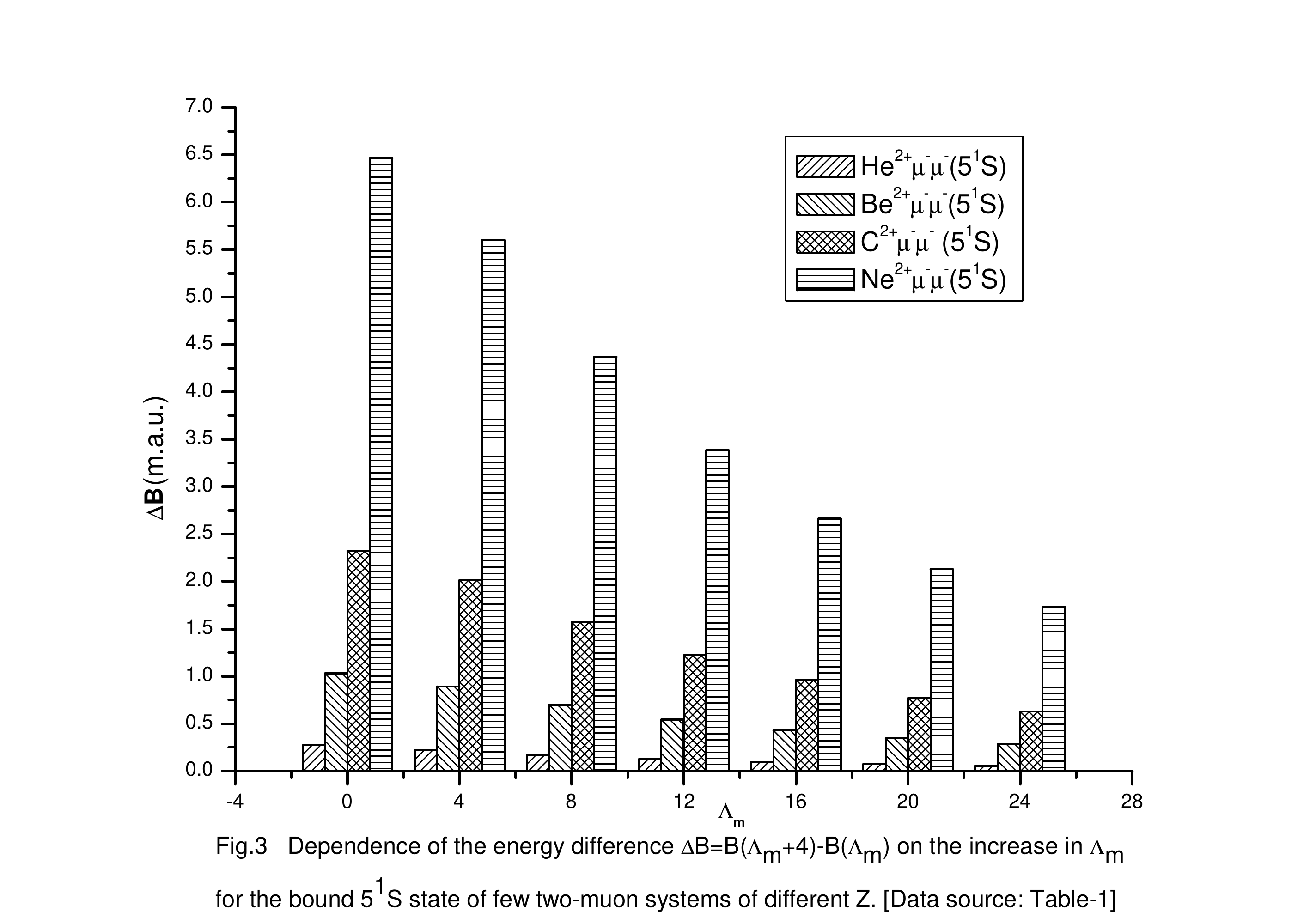}}
\caption{Dependence of the energy difference $\Delta B=B(\Lambda_m+4)-B(\Lambda_m)$ on the increase in $\Lambda_m$ for the bound 5$^1$S state of two-muon three-body systems with different nuclear charge Z.}
\label{fig:boxed_graphic}
\end{figure}
\newpage
\begin{figure}
\centering
\fbox{\includegraphics[width=0.75\linewidth, height=0.6\linewidth]{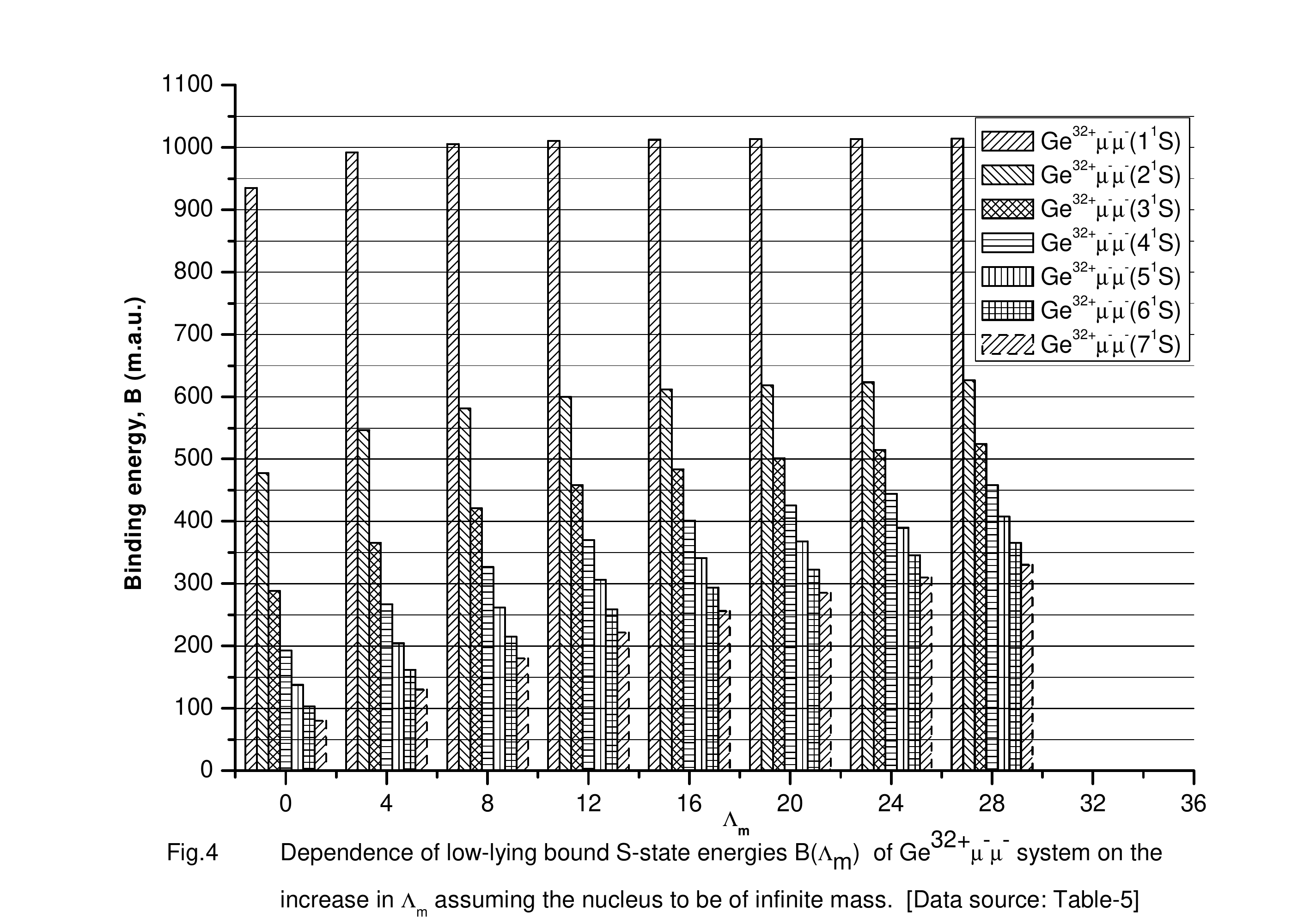}}
\caption{ Dependence of energy B$(\Lambda_m)$  of low-lying bound S-states of $^{\infty}$Ge$^{32+}\mu^-\mu^-$ three-body system on the increase in $\Lambda_m$.}
\label{fig:boxed_graphic}
\end{figure}
\newpage
\begin{figure}
\centering
\fbox{\includegraphics[width=0.75\linewidth, height=0.6\linewidth]{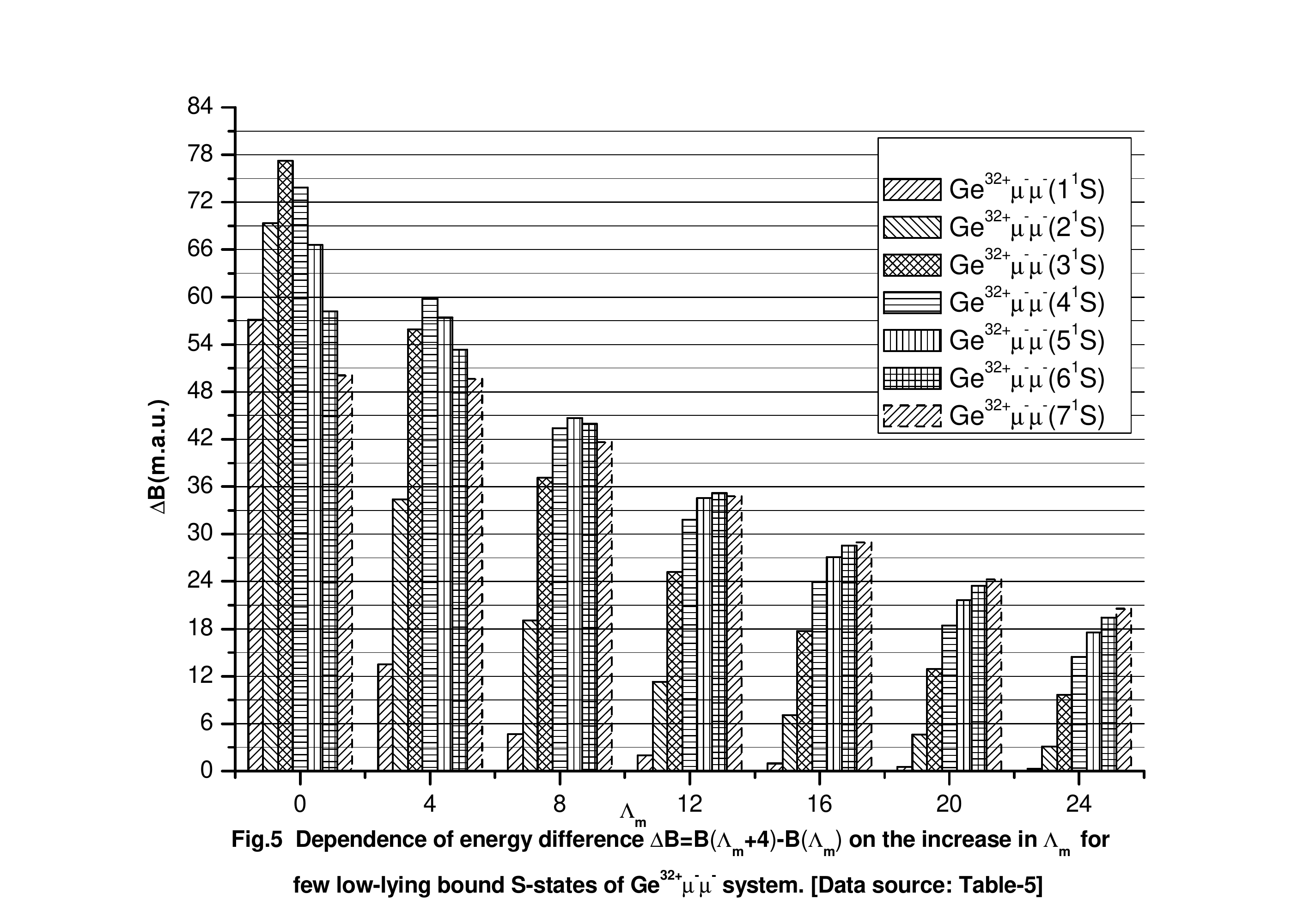}}
\caption{ Dependence of the energy difference $\Delta B=B(\Lambda_m+4)-B(\Lambda_m)$ on the increase in $\Lambda_m$ for the low-lying bound S-states of $^{\infty}$Ge$^{32+}\mu^-\mu^-$ three-body system.}
\label{fig:boxed_graphic}
\end{figure}
The calculated energies of the low-lying bound S-states, of  two-muon three-body systems of different nuclear charge Z (but of infinite nuclear mass), have been plotted against Z as shown in Fig. 2 to study the dependence of the bound state energies on the strength of the nuclear charge. The data used for Fig.2 is taken from column 3 of Table -3 and from column 7 of Table-4. And from Fig.2 it can be seen that the energy increases gradually with the increase in the strength nuclear charge Z. From the following empirical eq.(24, ) an estimate of the energy of the bound n$^1$S (n=1 to 6) state two-muon atom with a given Z can be done by an appropriate choice of value the set of parameters $\gamma_t (t=0,1,2,3)$ listed in Table-6
\begin{eqnarray}  
 B(Z)&=& \sum_{t=0}^3\gamma_tZ^t
 \end{eqnarray}
The values of the parameters $\gamma_t$ listed in Table-6 have been obtained by fitting the calculated energy data of Table-3 \& 4 for the bound n$^1$S (n=1 to 6) states of two-muon three-body systems with nucleus of different charge number (Z).
To see the effect of  nuclear charge strength on the convergence pattern of the energy of particular bound state we have plotted the difference in binding energies $\Delta B = B(\Lambda_m + 4) - B(\Lambda_m)$  against $\Lambda_m$  for the 5$^1$S state of few two-muon three-body systems with nucleus of different charge number (Z)  as a representative case using the calculated data presented  in Table-1. From Fig.3, it can be seen that the rate of convergence in energy with respect to increasing $\Lambda_m$  in the case of Ne$^{10+}\mu^-\mu^-$ having nuclear charge Z=10 is slower than that for He$^{2+}\mu^-\mu^-$ (Z=2) or Be$^{4+}\mu^-\mu^-$ (Z=4) or C$^{6+}\mu^-\mu^-$ (Z=4).  We have demonstrated the variations of- binding energy B($\Lambda_m$) with respect to increasing $\Lambda_m$ in Fig. 4 and the difference in energy $\Delta B$ against $\Lambda_m$ in Fig. 5 for few low-lying bound S-states of Ge$^{32+}\mu^-\mu^-$, as a representative case to study the pattern of convergence in energy of the low-lying bound S-states with respect to $\Lambda_m$  keeping the nuclear charge Z constant. 
By comparing, the relative change in height of the bars in Fig. 5, corresponding to different S-states for increasing $\Lambda_m$, it can be stated that  energy obtained for the lowest-lying bound S-states converges faster than the energy for higher excited S-states. In other words, energy of the 1$^1$S bound state tends towards convergence faster than that of the 2-7$^1$S bound states or energy of the 2$^1S$ bound state tends towards convergence better than that of the 3-7$^1$S states and so on. And finally, in Table-3, the energies of the low-lying bound S-states of several two-muon three-body systems calculated by an exact numerical solution of the coupled differential equation by the renormalized Numerov method [72] using RRC have been compared with the ones of the literature. 
\section{Conclusion}
\hspace*{1cm} In conclusion, we note that for systems with inter-particle interaction other than Coulomb or harmonic, use of RRC in HHE method becomes essential for the solution of the three-body Schr\H{o}dinger equation. Hence these coefficients are found to be of utmost importance for any type of interaction involved in three-body calculation. Further, the calculated energy for the low-lying  bound S-state at $\Lambda_m=28$ listed in column 3 of Table-3 in almost all cases are smaller than the corresponding values listed in column 5 \& 6 respectively. This is because of the eventual truncation of expansion basis to a maximum value of $\Lambda$ up to $\Lambda_m=28$ due to computer memory limitation. However, to get the solution at $\Lambda_m>28$, one may extrapolate the calculated energy values for $\Lambda_m=0, 4, 8,... ,28$ following the procedure described in [53]. The extrapolated energies for $\Lambda_m>28$ in steps of 4 have been presented in Table-2 for some representative cases. The extrapolated energy values (at $\Lambda_m=\Lambda_M$, a sufficiently large value) are listed in bold in column 4 of Table-3 and in columns 4 \& 8 of  Table-4 in bold. The extrapolated energies agree fairly with the corresponding exact values. For example, our calculated energy for the 1$^1$S bound state of $^{3}$H$^{1+}\mu^-\mu^-$ differs only by 0.1\% from those of Ancarani et al [1-2] and Frolov et al [73]. Again, since RRC's are independent of r, may be calculated once only and stored, resulting in an efficient and highly economical numerical computation. Finally, the present method being so simple, precise and highly effective for the description of two-muon three-body systems, can also be applied to more complex atomic and nuclear many-body systems.\\
\hspace*{1cm} The author acknowledges Aliah University for providing computer facilities.

\newpage
\section{Figure Caption}
\vspace{0.5cm}
\parskip 0.5cm
\parindent 0cm

Fig. 1. Choice of the Jacobi coordinates in the partition $\lq\lq i$" for a general three-body system. 

Fig. 2. Dependence of the energy, B($\Lambda_m=28$) of the low-lying bound S-states on the increase in nuclear charge Z.

Fig. 3. Dependence of the energy difference $\Delta B=B(\Lambda_m+4)-B(\Lambda_m)$ on the increase in $\Lambda_m$ for the bound 5$^1$S state of two-muon three-body  system with core of different nuclear charge Z.

Fig. 4. Dependence of the energy B($\Lambda_m$) on the increase in $\Lambda_m$ for few low-lying bound S-states of Ge$^{32+}\mu^-\mu^-$ (Z=32) three-body  system.

Fig. 5. Dependence of the energy difference $\Delta B=B(\Lambda_m+4)-B(\Lambda_m)$ on the increase in $\Lambda_m$ for few low-lying bound S-states of Ge$^{32+}\mu^-\mu^-$ (Z=32) three-body system.

\section{ Tables}
\begin{table}
\begin{center}
\begin{scriptsize}
{\bf Table-1. Pattern of convergence of energy calculated for the low-lying bound S-states in two-muon three-body systems with infinitely heavy nucleus for increasing $\Lambda_m$.}\\
\vspace{5pt}
\begin{tabular}{rllllllllllrl}\hline
 &\multicolumn{11}{c}{Bound state energy, B (=$B_{\Lambda_m}$ at $\Lambda=\Lambda_m$) in muon atomic unit (m.a.u.) in the $n^1S$ state of N$\mu^-\mu^-$ with nucleus N: }
\\
\cline{2-12}\\
$\Lambda_m$&  $^{\infty}$He$^{2+} _{1^1S}$&$^{\infty}$He$^{2+} _{5^1S}$   &$^{\infty}$Be$^{4+} _{1^1S}$&$^{\infty}$Be$^{4+} _{5^1S}$ & 
$^{\infty}$C$^{6+} _{1^1S}$&$^{\infty}$C$^{6+} _{5^1S}$& $^{\infty}$Ne$^{10+} _{1^1S}$&$^{\infty}$Ne$^{10+} _{5^1S}$ 
&$^{\infty}$Ge$^{32+} _{1^1S}$&$^{\infty}$Ge$^{32+} _{4^1S}$&$^{\infty}$Ge$^{32+} _{6^1S}$& \\\hline\hline
0	  &2.5000      &0.2992  & 12.2620  &1.8140  &29.4039     & 04.3497     & 85.8292   &12.6961 &0935.0070 &192.7639 &103.4429&\\
4	  &2.7844      &0.5783  & 13.2479  &2.8443  &31.5330     & 06.6726     & 91.5743   &19.1609 &0992.1403 &266.6499 &161.6256&\\
8   &2.8562      &0.8024  & 13.4821  &3.7361  &32.0419     & 08.6855     & 92.9513   &24.7617 &1005.6409 &326.4244 &214.9974&\\  	
12	&2.8760      &0.9710  & 13.5683  &4.4338  &32.2253     & 10.2576     & 93.4390   &29.1314 &1010.3062 &369.8567 &258.9279&\\  
16 	&2.8875      &1.0980  & 13.6056  &4.9774  &32.3041     & 11.4791     & 93.6479   &32.5196 &1012.2907 &401.6833 &294.1231&\\  
20  &2.8936      &1.1943  & 13.6245  &5.4071  &32.3437     & 12.4421     & 93.7519   &35.1850 &1013.2681 &425.5881 &322.6791&\\  
24  &2.8970      &1.2681 	& 13.6356  &5.7526  &32.3653     & 13.2143     & 93.8086   &37.3179 &1013.7997 &443.9859 &346.1072&\\  
28  &2.8990      &1.3250  & 13.6418  &6.0349  &32.3781     & 13.8437     & 93.8422   &39.0526 &1014.1128 &458.4522 &365.5627&\\  \\ \hline\hline
\end{tabular}
\end{scriptsize}
\end{center}
\end{table}

\newpage
\begin{table}
\begin{center}
\begin{scriptsize}
{\bf Table-2.   Pattern of convergence of energy calculated for the low-lying bound S-states in two-muon three-body systems with infinitely heavy nucleus for increasing $\Lambda_m$.}\\
\vspace{5pt}
\begin{tabular}{rllllllllllrl}\hline
 &\multicolumn{11}{c}{BE (=$B_{\Lambda_m}$ at $\Lambda=\Lambda_m$) in muon atomic unit (m.a.u.) in the $n^1S$ state of N$\mu^-\mu^-$ with core N: }
\\
\cline{2-12}\\
$\Lambda_m$&  $^{\infty}$He$^{2+} _{1^1S}$&$^{\infty}$He$^{2+} _{5^1S}$   &$^{\infty}$Be$^{4+} _{1^1S}$&$^{\infty}$Be$^{4+} _{5^1S}$ & 
$^{\infty}$C$^{6+} _{1^1S}$&$^{\infty}$C$^{6+} _{5^1S}$& $^{\infty}$Ne$^{10+} _{1^1S}$&$^{\infty}$Ne$^{10+} _{5^1S}$ 
&$^{\infty}$Ge$^{32+} _{1^1S}$&$^{\infty}$Ge$^{32+} _{4^1S}$&$^{\infty}$Ge$^{32+} _{6^1S}$& \\\hline\hline
32 & 2.9003 & 1.3719 & 13.6458 & 6.2738  &32.3865  & 14.3759 & 93.8640 & 40.5195&1014.3157&470.2364&382.5157  \\  
36 & 2.9011 & 1.4109 & 13.6486 & 6.4736  &32.3920 & 14.8208 & 93.8785  & 41.7442&1014.4502&479.7773& 396.9926    \\  
40 & 2.9017 & 1.4437 & 13.6504 & 6.6422  &32.3958  & 15.1956 & 93.8884 & 42.7748&1014.5429&487.5843& 409.4294  \\  
44 & 2.9021 & 1.4715 & 13.6518 & 6.7854  &32.3985  & 15.5135 & 93.8956  & 43.6483&1014.6088&494.0339&420.1730 \\  
48 & 2.9024 & 1.4951 & 13.6528 & 6.9077 &32.4005 & 15.7849 & 93.9008  & 44.3934&1014.6569&499.4087& 429.5022    \\  
52 & 2.9026 & 1.5154 & 13.6535 & 7.0129  &32.4020  & 16.0180 & 93.9047 & 45.0330 &1014.6929 &503.9236& 437.6421 \\  
56 & 2.9028 & 1.5330 & 13.6541 & 7.1038  &32.4032  & 16.2194 & 93.9076 & 45.5845 &1014.7204 &507.7438& 444.7765    \\  
60 & 2.9029 & 1.5481 & 13.6545 & 7.1829  &32.4040  & 16.3942 & 93.9099  & 46.0639 &1014.7416 &510.9981& 451.0558    \\  
64 & 2.9031 & 1.5613 & 13.6548 & 7.2519  &32.4047  & 16.5467 & 93.9118  & 46.4815 &1014.7584 &513.7873& 456.6044    \\ 
68 & 2.9031 &1.5728 & 13.6551 & 7.3124 &32.4053 & 16.6804 & 93.9132&
46.8473&1014.7718&516.1917&  461.5256   \\  
72 & 2.9032 &1.5830 & 13.6553 & 7.3657 &32.4057 &16.7981 &93.9144 &47.1690     &1014.7826&518.2755&465.9057\\  
76 &2.9033 & 1.5919 & 13.6555 &7.4128  &32.4061 &16.9020 &93.9153 &47.4531   &1014.7914&520.0903&469.8169\\  
80 &2.9033 & 1.5999 & 13.6557 & 7.4546 &32.4064 & 16.9942 & 93.9161 &47.7049 &1014.7987 &521.6781& 473.3204    \\ 
84 &2.9034 & 1.6069 & 13.6558 & 7.4918 &32.4067 &17.0762 & 93.9168 & 47.9289 &1014.8047 &523.0735& 476.4679 \\  
88 & 2.9034 & 1.6132 &13.6559 & 7.5251 &32.4069 &17.1495 & 93.9173     & 48.1289 &1014.8098 &524.305& 479.3036  \\  
92& 2.9034 & 1.6189 &13.6560 &7.5549 &32.4071 & 17.2151 & 93.9178 &48.3079    &1014.8141 &525.3949 &481.8651\\   
96& 2.9034 & 1.6239 &13.6561 &7.5817 &32.4072 &17.2740 & 93.9182 &48.4686   &1014.8177 &526.3640 &484.1848\\   
100 & 2.9035 & 1.6285 & 13.6561 &7.6058 &32.4073 & 17.3271 &93.9185 & 48.6134 &1014.8209 &527.2283 &486.2905   \\   
104 &2.9035  &1.6326 & 13.6562 & 7.6276 &32.4075 & 17.3751  & 93.9188 & 48.7441 &1014.8236 &528.0016&488.2063    \\   
108 &2.9035 & 1.6363 & 13.6562 & 7.6474 &32.4076  &17.4185 & 93.9191 & 48.8624 &1014.8259&528.6955&489.9532   \\   
112& 2.9035 & 1.6396 & 13.6563 & 7.6654 & 32.4076 & 17.4579& 93.9193 & 48.9698 &1014.8279&529.3201&491.5494    \\   
116 &2.9035 & 1.6427 & 13.6563 &7.6817 & 32.4077  &17.4938 & 93.9195   & 49.0675&1014.8297&529.8836& 493.0109   \\   
120 &2.9035 & 1.6455  &13.6564  &7.6966 &32.4078 &17.5265 & 93.9197 & 49.1565& 1014.8312&530.3934& 494.3514   \\   
124 & 2.9035  & 1.6480 &13.6564 &7.7102 & 32.4078 & 17.5563 &93.9198 &	49.2378 &1014.8326 &530.8558& 495.5833     \\   
128 & 2.9035  & 1.6504 & 13.6564 &7.7227 &32.4079 & 17.5837 &93.9200&	 49.3122&1014.8338&531.2762& 496.7174   \\
132 &2.9036  & 1.6525 & 13.6564 &7.7341 &32.4079 & 17.6088 &93.9201 &49.3804 &1014.8349 &531.6591 &497.7632 \\
136 & 2.9036 & 1.6545 &13.6564  & 7.7446 &32.4080 & 17.6318 & 93.9202 &49.4431 &1014.8359 &532.0089& 498.7290   \\
140 & 2.9036  & 1.6563 & 13.6565& 7.7543 & 32.4801 &17.6531 & 93.9203 & 49.5008 &1014.8367&532.3286&499.6225   \\
144 &  2.9036  & 1.6579  &13.6565  & 7.7633& 32.4080  &17.6726 & 93.9204     & 49.5540&1014.8375&532.6218&  500.4502    \\
148 & 2.9036 &1.6595 &13.6565 &7.7715 &32.4081 &17.6907 &93.9204&     49.6031 &1014.8382 &532.8911 &501.2181 \\      
152 & 2.9036&  1.6609&  13.6565&   7.7791 & 32.4081 & 17.7074 &    93.9205 &49.6486 &1014.8388 &533.1389&501.9315\\     
156 & 2.9036& 1.6622& 13.6565& 7.7862 & 32.4081 & 17.7229 & 93.9206& 49.6906 &1014.8394 &533.3670 &502.5952\\    
160 & 2.9036& 1.6634& 13.6565& 7.7928 & 32.4081 & 17.7373 &  93.9206& 49.7297 &1014.8399 &533.5777& 503.2134\\     
164 & 2.9036& 1.6646& 13.6565& 7.7989 & 32.4082 & 17.7506 & 93.9207 & 49.7659 &1014.8404 & 533.7725 &503.7899\\     
168 & 2.9036& 1.6656  &13.6566 & 7.8046 & 32.4082 & 17.7630& 93.9207& 49.7996 &1014.8408 & 533.9528 &504.3282\\  
172 & 2.9036& 1.6666& 13.6566&   7.8099 & 32.4082 & 17.7746 &    93.9208& 49.8310 &1014.8412 &534.1200 &504.8314\\   
176 & 2.9036& 1.6675& 13.6566&   7.8148 & 32.4082 &    17.7854 &    93.9208&     49.8603 &1014.8415 &534.2753 &505.3024\\   
180 & 2.9036& 1.6684& 13.6566& 7.8194 & 32.4082 & 17.7955 & 93.9208 & 49.8876 &1014.8418 &534.4197 &505.7436\\    
184 & 2.9036& 1.6692& 13.6566& 7.8237 & 32.4082&  17.8049 &    93.9209 & 49.9132 &1014.8421 &534.5541&506.1573\\ 
   188 &  2.9036 & 1.6699& 13.6566& 7.8277 & 32.4082 & 17.8137 & 93.9209 &     49.9371 &1014.8424 &534.6794&506.5458\\   
   192 & 2.9036&  1.6706 & 13.6566 & 7.8315 & 32.4083 & 17.8219 &    93.9209 &49.9595 &1014.8427&534.7963&506.9109\\    
   196 & 2.9036&    1.6713& 13.6566& 7.8351 & 32.4083 & 17.8297 &    93.9209 &49.9804 &1014.8429 &534.9056 &507.2542\\   
   200 & 2.9036& 1.6719& 13.6566& 7.8384 & 32.4083 & 17.8369 & 93.9210& 50.0001 &1014.8431 &535.0078&507.5775\\   
   204 & 2.9036& 1.6724& 13.6566& 7.8415 & 32.4083 & 17.8438 & 93.9210& 50.0187 &1014.8433 &535.1035&507.8821\\     
   208 & 2.9036& 1.6730& 13.6566& 7.8444 & 32.4083 & 17.8502 & 93.9210& 50.0361 &1014.8435 &535.1932&508.1694\\      
   212 & 2.9036& 1.6735&  13.6566& 7.8472 & 32.4083 & 17.8562 & 93.9210& 50.0525 &1014.8437&535.2774&508.4406\\     
   216 & 2.9036& 1.6740& 13.6566& 7.8498 & 32.4083 & 17.8619 & 93.9210&  50.0679 &1014.8438 &535.3565&508.6968\\       
   220 & 2.9036&  1.6744& 13.6566& 7.8523 & 32.4083 & 17.8673 & 93.9211& 50.0825 &1014.8440 &535.4309&508.9391\\ \hline\hline
\end{tabular}
\end{scriptsize}
\end{center}
\end{table}

\newpage
\begin{table}
\begin{center}
\begin{scriptsize}
{\bf Table-3. Comparison of calculated energy for the low-lying bound S-states of two-muon three-body systems with the ones of the  literature.}\\
\vspace{5pt}
\begin{tabular}{llllcc}\hline
System &State&\multicolumn{4}{c}{Bound state energies expressed in muon atomic unit (m.a.u.)}
\\
\cline{3-6}
&&\multicolumn{2}{c}{\underline{Present Calculation}}&\multicolumn{2}{c}{Other Results}\\
\cline{5-6}
&& $B_{\Lambda_m=28}$ & $B_{\Lambda_m=\Lambda_M}$ &Ref[1-2]&Ref[73]\\\hline
$^1$H$^+\mu^-\mu^-$&$1^1S$&0.46735477&{\bf 0.46998956} &{\it 0.47093683}& {\it 0.47186663}\\
&$2^1S$&0.35545924&{\bf 0.37119584}& - &-\\
&$3^1S$&0.20247120&{\bf 0.21630250 }&-&- \\
&$4^1S$&0.01665469&{\bf 0.02645864}&-&- \\\\

$^2$H$^+\mu^-\mu^-$&$1^1S$&0.49390510&{\bf 0.49657102} &{\it 0.49718941}&{\it 0.49810300}\\
&$2^1S$&0.38354583&{\bf 0.40146950}& -  &-\\
&$3^1S$&0.23613474&{\bf 0.25209024}& -  &- \\
&$4^1S$&0.04930593&{\bf 0.06594398}& -  &- \\\\

$^3$H$^+\mu^-\mu^-$&$1^1S$&0.50344776&{\bf 0.50613881} &{\it 0.50663960}&{\it 0.50754448}\\
&$2^1S$&0.39345193&{\bf 0.41232457}& -  &-\\
&$3^1S$&0.24810205&{\bf 0.26501518}& -  &-\\
&$4^1S$&0.06046861&{\bf 0.07270507}& -  &-\\\\

$^{\infty}$H$^+\mu^-\mu^-$&$1^1S$& 0.52379740&{\bf 0.52656473}&{\it 0.52686030}&\\ 
&$2^1S$&0.41402136&{\bf 0.43528652}&- &-\\
&$3^1S$&0.27315517&{\bf 0.29245473}&- &-\\
&$4^1S$&0.08661911&{\bf 0.09985389}&- &-\\\\

$^3$He$^{2+}\mu^-\mu^-$&$1^1S$&2.78878545&{\bf 2.79346866}&{\it 2.79024400} &\\
& $2^1S$& 1.97550223&{\bf 2.05919044}&{\it 2.06728676}
&-\\ 
& $3^1S$& 1.71968500&{\bf 1.9312509}&{\it 1.98620901}
&-\\ 
& $4^1S$& 1.50294490&{\bf 1.79815522}&{\it 1.95287356}
&-\\ 
& $5^1S$& 1.24794925&{\bf 1.54097190}&-&-\\\\

$^4$He$^{2+}\mu^-\mu^-$&$1^1S$& 2.81508427&{\bf 2.81978459}&{\it 2.81668000}&-\\
&$2^1S$& 1.99383842&{\bf 2.07829754}&{\it 2.08596390}
 &-\\
&$3^1S$& 1.73631922&{\bf 1.95020173}&{\it 2.00412688}
&-\\
&$4^1S$& 1.51970869&{\bf 1.82076365}&{\it 1.97051576}
 &-\\
&$5^1S$& 1.26703166&{\bf 1.62458960}&- &-\\\\

$^{\infty}$He$^{2+}\mu^-\mu^-$&$1^1S$&2.89900954&{\bf 2.90358186}&{\it 2.90107355}&-\\
&$2^1S$&2.05144929&{\bf 2.14605623}&{\it 2.14532259}
&-\\
&$3^1S$&1.78764955&{\bf 2.00890603}&{\it 2.06098528}
&-\\
&$4^1S$&1.57062240&{\bf 1.89019948}&{\it 2.02652301}
&-\\
&$5^1S$&1.32504206&{\bf 1.68224817}&- &-\\
\\

$^6$Li$^{3+}\mu^-\mu^-$&$1^1S$&7.13185573&{\bf 7.14063441}&{\it 7.13629800}&-\\
&$2^1S$&4.79423674&{\bf 4.94591157 }&{\it 4.94552512}
&-\\
&$3^1S$&4.13413530&{\bf 4.58363159 }&{\it 4.64520134}
&-\\
&$4^1S$&3.65339921&{\bf 4.39998451 }&{\it 4.53915464}
&-\\
&$5^1S$&3.26473644&{\bf  3.66296332}& -&-\\\\

$^7$Li$^{3+}\mu^-\mu^-$&$1^1S$&7.15138919&{\bf 7.16018343}&{\it 7.15593300}&-\\
&$2^1S$&4.80710867 &{\bf 4.95919841 }&{\it 4.95863104}
&-\\
&$3^1S$&4.14544926 &{\bf 4.59621085 }&{\it 4.65748995}
&-\\
&$4^1S$&3.66356880 &{\bf 4.41233128 }&{\it 4.55118210}
&-\\
&$5^1S$&3.27399613 &{\bf 4.27540046}&-&-\\\\

$^{\infty}$Li$^{3+}\mu^-\mu^-$&$1^1S$&7.27137265&{\bf 7.28028376}&{\it 7.27659955}&-\\ 
&$2^1S$&4.88564379 &{\bf 5.04036937}& {\it 5.03920100}
&-\\
&$3^1S$&4.21376895 &{\bf 4.67227404}&{\it 4.73289975}
&-\\
&$4^1S$&3.72445246 &{\bf 4.48616767}&{\it 4.62496213}
&-\\
&$5^1S$&3.32900484 &{\bf 4.39042199}&-&-\\\hline
\end{tabular}
\end{scriptsize}
\end{center}
\end{table}

\newpage
\begin{table}
\begin{center}
\begin{scriptsize}
{\bf Table-4. Calculated energy of the low-lying bound S-states  of two-muon three-body systems where reference values are not available .}\\
\vspace{5pt}
\begin{tabular}{llrr|llrr}\hline
\multicolumn{8}{c}{BE (=$B_{\Lambda_m}$ at $\Lambda=\Lambda_m$) in muon atomic unit (m.a.u.)}
\\
\cline{1-8}
System&State& $B_{\Lambda_m=28}$ & $B_{\Lambda_m=\Lambda_M}$&System&State&$B_{\Lambda_m=28}$ & $B_{\Lambda_m=\Lambda_M}$\\\hline
$^{9}$Be$^{4+}\mu^-\mu^-$&$1^1S$&13.46836088&{\bf13.48275321}&$^{\infty}$Be$^{4+}\mu^-\mu^-$& $1^1S$&13.64177142&{\bf13.65664476}\\
&$2^1S$&8.83821750 &{\bf  }&&$2^1S$&8.94960222 &{\bf 9.19245339}\\
&$3^1S$&7.57239016 &{\bf }&&$3^1S$&7.66846380  &{\bf 8.44750096}\\
&$4^1S$&6.67577395 &{\bf }&&$4^1S$&6.76103019  &{\bf 8.08196767}\\
&$5^1S$&5.95840820 &{\bf  }&&$5^1S$&6.03494406 &{\bf 7.89539386}\\
&$6^1S$&5.36216959 &{\bf  }&&$6^1S$&5.43140179 &{\bf 7.83720103}\\

$^{10}$B$^{5+}\mu^-\mu^-$&$1^1S$&21.75983645&{\bf21.78127757}&$^{\infty}$B$^{5+}\mu^-\mu^-$& $1^1S$&22.01061306&{\bf22.03226776}\\
& $2^1S$&14.08507649 &{\bf }&&$2^1S$&14.24458102 &{\bf 14.59474446}\\
& $3^1S$&12.01556673 &{\bf }&&$3^1S$&12.15246198 &{\bf 13.33446396}\\
& $4^1S$&10.57611993 &{\bf }&&$4^1S$&10.69733572 &{\bf 12.73078665}\\
& $5^1S$& 9.43102804 &{\bf }&&$5^1S$& 9.53968706 &{\bf 12.41799056} \\
& $6^1S$& 8.48198833 &{\bf }&&$6^1S$& 8.58016529 &{\bf 12.31099898} \\

$^{12}$C$^{6+}\mu^-\mu^-$&$1^1S$&32.07123946&{\bf32.10120848}&$^{\infty}$C$^{6+}\mu^-\mu^-$&$1^1S$&32.37814008&{\bf32.40835778}\\
&$2^1S$&20.57703042 &{\bf  }&&$2^1S$&20.77100359 &{\bf 21.24914880}\\
&$3^1S$&17.50000505 &{\bf  }&&$3^1S$&17.66580282 &{\bf 19.33526864}\\
&$4^1S$&15.38686803 &{\bf  }&&$4^1S$&15.53337783 &{\bf 18.43190163}\\
&$5^1S$&13.71251849 &{\bf  }&&$5^1S$&13.84366250 &{\bf 17.96089243}\\
&$6^1S$&12.32751704 &{\bf  }&&$6^1S$&12.44587479 &{\bf 17.78810696}\\

$^{16}$O$^{8+}\mu^-\mu^-$&$1^1S$&58.69106610&{\bf58.74242738}&$^{\infty}$O$^{8+}\mu^-\mu^-$&$1^1S$&59.11039341&{\bf59.16209025}\\
&$2^1S$&37.25593866&{\bf  }&&$2^1S$&37.51901718 &{\bf 38.31361890}\\
&$3^1S$&31.55703793&{\bf  }&&$3^1S$&31.78070668 &{\bf 34.67637674}\\
&$4^1S$&27.70765768&{\bf  }&&$4^1S$&27.90478889 &{\bf 32.99157418}\\
&$5^1S$&24.67328384&{\bf  }&&$5^1S$&24.84941473 &{\bf 32.10746357}\\
&$6^1S$&22.16951745&{\bf  }&&$5^1S$&22.32824372 &{\bf 31.77966747}\\

$^{20}$Ne$^{10+}\mu^-\mu^-$&$1^1S$&93.31027811&{\bf93.38881160}&$^{\infty}$Ne$^{10+}\mu^-\mu^-$&$1^1S$&93.84221663&{\bf93.92117897}\\
&$2^1S$&58.86289409&{\bf  }&&$2^1S$&59.19521811&{\bf 60.38674068}\\
&$3^1S$&49.73219582&{\bf  }&&$3^1S$&50.01379450&{\bf 54.46665222}\\
&$4^1S$&43.62792071&{\bf  }&&$4^1S$&43.87570707&{\bf 51.76123767}\\
&$5^1S$&38.83143483&{\bf  }&&$5^1S$&39.05257343&{\bf 50.33501851}\\
&$6^1S$&34.87973178&{\bf  }&&$6^1S$&35.07883941&{\bf 49.78428221}\\

$^{28}$Si$^{14+}\mu^-\mu^-$&$1^1S$&186.58181660&{\bf186.73197776}&$^{\infty}$Si$^{14+}\mu^-\mu^-$&$1^1S$&187.33991986&{\bf187.49065697}\\
&$2^1S$&116.87206982 &{\bf  }&&$2^1S$&117.34327599 &{\bf 119.56801338}\\
&$3^1S$& 98.44274212 &{\bf  }&&$3^1S$& 98.84039229 &{\bf 107.42053496}\\
&$4^1S$& 86.27128009 &{\bf  }&&$4^1S$& 86.62050862 &{\bf 101.93205784}\\
&$5^1S$& 76.74351154 &{\bf  }&&$5^1S$& 77.05476512 &{\bf 99.03362906}\\
&$6^1S$& 68.90786452 &{\bf  }&&$6^1S$& 69.18781295 &{\bf 97.87364713}\\

$^{40}$Ar$^{18+}\mu^-\mu^-$&$1^1S$&312.12923132&{\bf312.37368777}&$^{\infty}$Ar$^{18+}\mu^-\mu^-$&$1^1S$&313.01778694&{\bf 313.26290079}\\
&$2^1S$&194.71002127 &{\bf  }&&$2^1S$&195.26005541 &{\bf 198.82967402}\\
&$3^1S$&163.70716920 &{\bf  }&&$3^1S$&164.16997440 &{\bf 178.19874736}\\
&$4^1S$&143.38039937 &{\bf  }&&$4^1S$&143.78629491 &{\bf 168.95116040}\\
&$5^1S$&127.50400701 &{\bf  }&&$5^1S$&127.86542065 &{\bf 164.04954287}\\
&$6^1S$&114.46081978 &{\bf  }&&$6^1S$&114.78563676 &{\bf162.06610437}\\

$^{73}$Ge$^{32+}\mu^-\mu^-$&$1^1S$&1012.48293571&{\bf 1013.21422357}&$^{\infty}$Ge$^{32+}\mu^-\mu^-$
&$1^1S$&1014.11284946 &{\bf 1014.84508035}\\
&$2^1S$&625.16157278 &{\bf  }&&$2^1S$&626.14680929 &{\bf 636.45151611}\\
&$3^1S$&523.30955758 &{\bf  }&&$3^1S$&524.12890560&{\bf 567.25338865}\\
&$4^1S$&457.73615497 &{\bf  }&&$4^1S$&458.45220105 &{\bf  537.01104413}\\
&$5^1S$&406.77091860 &{\bf  }&&$5^1S$&407.40717276 &{\bf  521.09516457}\\
&$6^1S$&364.99173275 &{\bf  }&&$6^1S$&365.56269120 &{\bf  514.59886056}\\

$^{132}$Xe$^{54+}\mu^-\mu^-$&$1^1S$&3031.51129859&{\bf 3033.28646725}&$^{\infty}$Xe$^{54+}\mu^-\mu^-$
&$1^1S$&3034.34403549 &{\bf 3036.12024452}\\
&$2^1S$&1834.09223260 &{\bf  }&&$2^1S$&1835.74140513 &{\bf 1858.28223928}\\
&$3^1S$&1518.77351695 &{\bf  }&&$3^1S$&1520.11029918 &{\bf 1631.62069230}\\
&$4^1S$&1325.18609219 &{\bf  }&&$4^1S$&1326.34795821 &{\bf 1537.18457058}\\
&$5^1S$&1176.27706247 &{\bf  }&&$5^1S$&1177.30693158 &{\bf 1486.01523381}\\
&$6^1S$&1054.70453313 &{\bf  }&&$6^1S$&1055.62738998 &{\bf 1432.37654976}\\
&$7^1S$& 952.71659241 &{\bf  }&&$7^1S$& 953.55000451 &{\bf 1365.70825884}\\\hline
\end{tabular}
\end{scriptsize}
\end{center}
\end{table}

\begin{table}
\begin{center}
\begin{scriptsize}
{\bf Table-5. Pattern of convergence of energy calculated for the low-lying bound S-states of $^{\infty}$Ge $\mu^-\mu^-$  for increasing $\Lambda_m$.}\\
\vspace{5pt}
\begin{tabular}{cccccccc}\hline
 &\multicolumn{7}{c}{Calculated energy, B(=$B_{\Lambda_m}$ at $\Lambda=\Lambda_m$) (in muon atomic unit,  m.a.u.) for the bound $n^1S$ state of $^{\infty}$Ge $\mu^-\mu^-$ system.}\\ \cline{2-8}\\
$\Lambda_m$   &B($1^1S$)  &$B{2^1S}$  &B($3^1S$)   &B($4^1S$)   &B($5^1S$)   &B($6^1S$) &B($7^1S$)\\\hline\hline
0	& 935.007025     &477.249742     &288.344360   &192.763931  &137.853353  &103.442913  & 80.470732\\
4	& 992.140293	   &546.578558    &365.564398   &266.649941  &204.450067  &161.625563  &130.571487\\
8	&1005.640886   &580.955261    &421.434876   &326.424415  &261.854693  &214.997432   &180.198265\\
12	&1010.306245   &600.025927    &458.569924   &369.856728  &306.547289  &258.927922   &221.837017\\
16	&1012.290752   &611.321603    &483.781416   &401.683338  &341.111126  &294.123100   &256.612742\\
20	&1013.268082   &618.390002    &501.538712   &425.588085  &368.222059  &322.679110   &285.537726\\
24	&1013.799689   &623.011450    &514.462456   &443.985939  &389.857115  &346.107219   &309.821712\\
28	&1014.112849	   &626.146809   &524.128906   &458.452201   &407.407173  &365.562691   &330.379238 \\ \\ \hline\hline

\end{tabular}
\end{scriptsize}
\end{center}
\end{table}

\begin{table}
\begin{center}
\begin{scriptsize}
{\bf Table-6. Values of parameters involved in eq(25) obtained by best fit of calculated energies.}\\
\vspace{5pt}
\begin{tabular}{lrrrr}\hline
State&\multicolumn{4}{c}{Parameters (in muon atomic unit, 1mau=27.12$m_{\mu}$eV=5.6076 KeV)}
\\
\cline{2-5}
&$\gamma_0$ & $\gamma_1$ & $\gamma_2$ &$\gamma_3$ \\\hline
1$^1$S&-0.05809&-0.48651&0.9783&8.52919E-4\\
2$^1$S&-0.0091&-0.19528&0.60854&2.82456E-4\\
3$^1$S&-0.09375&-0.10071&0.5094&1.77743E-4\\
4$^1$S&-0.20682&-0.0404&0.44331&1.83214E-4\\
5$^1$S&-0.12444&-0.04807&0.39539&1.27868E-4\\
6$^1$S&-0.0575&-0.05354&0.3558&9.15059E-5\\
\hline
\end{tabular}
\end{scriptsize}
\end{center}
\end{table}

\end{document}